\documentclass[a4paper,10pt]{article}

\usepackage{times}
\usepackage[noadjust]{cite}
\usepackage{enlarge}

\usepackage{tex_andy}
\usepackage{equations_andy}
\usepackage{theorems_andy}
\usepackage{epic,eepic}
\usepackage{epsfig}
\usepackage{sections_andy}
\usepackage{figures_andy}
\usepackage{algorithms_andy}
\makeatletter
\@ndyTheorem{Invar}{Invariance}{Invariance}
\setcommand{\CommRight}[1]{%
  \mbox{}\hfill{}/* {\normalfont\slshape\selectfont #1} */}
\makeatother
\setcommand{\Intvl}[2]{\:[#1, #2)\:}

\title{
  Data Structure for a Time-Based Bandwidth Reservations Problem
}

\author{
  Andrej Brodnik\thanks{
    Department of Theoretical Computer Science, Institute of
    Mathematics, Physics, and Mechanics, Ljubljana, Slovenia} $\ ^\dag$
  \and
  Andreas Nilsson\thanks{
    Department of Computer Science and Electrical Engineering, Lule{\aa}
    University of Technology, Lule{\aa}, Sweden}
}

\begin{document} 

\maketitle

\begin{abstract}
  We discuss a problem of handling resource reservations. The resource
  can be reserved for some time, it can be freed or it can be queried
  what is the largest amount of reserved resource during a time
  interval. We show that the problem has a lower bound of
  $\Omega(\log n)$ per operation on average and we give a matching
  upper bound algorithm. Our solution also solves a dynamic version of
  the related problems of a prefix sum and a partial sum.
\end{abstract}

\section{Introduction} 

In Computer Communications we need to make bandwidth reservations
over the Internet to provide \emph{Quality of Service} (\emph{QoS})
for the end users. The IETF (Internet Engineering Task Force) defined
a standard for Integrated Services in routers (\cite{rfc2211,rfc2212})
and the end-to-end reservation setup protocol RSVP (\cite{rfc2205}).
Since the protocol does not scale well (\cite{rfc2208})
IETF came up with a new approach, known as
\emph{differentiated services} (\emph{diffserv},
\cite{diffserv-arch}).
Schel\'{e}n et.\ al.\
(\cite{Schelen9705:Sharing,Schelen9711:Agent}) used the
\emph{diffserv} to design a new QoS architecture. In this architecture
they provide virtual leased lines using the differentiated services to
perform admission control through the system of agents. The agents
work on per-hop basis and they need to maintain a database of the
reservations made on their hop. In the backbone of the Internet it
will most likely be many reservations to administrate and hence the
use of an efficient data structure will be required. Moreover, in the
design of the agents the authors propose that a single agent
administrates several hops to make it more attractive for the ISP's
(Internet Service Provider). Such a scenario even increases the need
for use of an efficient data structure.
Therefore Schel\'{e}n et.\ al.\ in \cite{schelen99} proposed a
solution that was, however, limited to a predefined set of possible
time intervals over which the reservation could be made.

The bandwidth reservation problem is a special case of a more generic
problem, where we need to administrate a limited resource over the time;
e.g.\ use of human resources, computational power of super-computer, pool of
cars etc. Although the solution in this paper covers all these problems, we
use the term \emph{bandwidth} when we talk about the reserved resource.

\bDefn{BRP}
In the \emph{bandwidth reservation problem} we have a fixed amount of
bandwidth to administer. Customers want to make reservations
$R=\{B,I\}$ for a part of the bandwidth $B$ during a time interval
$I=\Intvl{t_0}{t_1}$ ($t_0 < t_1$, the interval starts at time $t_0$
and ends at time $t_1$, and it includes $t_0$). The operations to
support, besides initialization and destruction, are:
\begin{itemize}
\item  $\mathtt{Reserve}(B,I)$, that reserves $B$ units of bandwidth
  for the time period $I=\Intvl{t_0}{t_1}$, where $t_0\leq t_1$.
\item $\mathtt{Free}(B,I)$, that frees the reserved bandwidth $B$
  during the interval $I$. Note that freeing the bandwidth is the same
  as making a reservation with a negative bandwidth.
\item $\mathtt{MaxReserved}(I)$, that returns maximum reserved
  bandwidth during the interval $I$.
\end{itemize}
\eDefn

For the sake of clarity, we sometimes use the subscripts $q$ and $r$
for queries and reservations respectively. For example, a reservation
interval $I_r = \Intvl{t_{r0}}{t_{r1}}$. In the paper we also use the
notation $\max(x, y)$ denoting a function returning the bigger of $x$
and $y$.

\subsection{Literature background}

In the literature we could not find any reference to the bandwidth
reservation problem with an arbitrary reservation interval -- i.e.\
interval where endpoints are not drawn from a predefined set.
However, the problem is similar to problems we find in other fields of
computer science that handle intervals on a real line (e.g.\
computational geometry, dynamic computation and geometric search
\cite{mehlhorn86routing,arge96buffer,
  CDR-CGPGS-99,eppstein91dynamic}). These problems are
generally solved using \emph{segment trees}
(\cite{preparata85computational,Meh84a}), which were introduced by
Bentley (\cite{bentley77algorithms}) as a solution to the Klee's
rectangle problem (\cite{k-cmbbc-77}).
The limitation in all these problems is that the end-points of the
intervals belong to a \emph{fixed set} of points. In our problem we
have no such a set.

Kuchem et.\ al.\ in (\cite{ALGOR::KuchemWW1996}) presented in a way
similar data structure to ours, although it still deals with a fixed
set of points.  They use the structure in a VLSI design. Bose et.\
al.\ independently developed in \cite{BoKreMaMoMo2001} a similar
data structure to solve a number of geometric problems.

Another pair of related problems are the well studied partial sum
problem (\cite{Fredman82}, brief in \cite{HusRau-ICALP-98}), and the
prefix sum problem (\cite{Fredman82}). In the prefix sum problem we
have an array $V(i), 1 \leq i \leq n$ on which we want to perform
these two operations: (1) $\mathtt{Update}(i,x)$: $V(i)=V(i)+x$; and
(2) $\mathtt{Retrieve}(m)$: $\sum_{k=1}^mV(k)$ for arbitrary values of
$i$, $x$ and $m$. In \cite {Fredman82} Fredman shows a lower bound of
$\Omega(\log n)$ for the problem under the comparison based model. In
the same paper Fredman also presents an algorithm with a matching
upper bound.

In the rest of the paper we first show that the logarithmic lower
bound carries over to the bandwidth reservation problem. We continue
with a presentation of a data structure we call \emph{BinSeT} (binary
segment tree) that gives us a matching upper bound. We conclude the
paper with final remarks.

\section{Lower bound} 

\bTheo{LB}
Given an arbitrary sequence of operations from a bandwidth reservation
problem, each of them requires at least
$\Omega (\log n)$ comparisons on the average, where $n$ is the number
of intervals we are dealing with.
\eTheo

\bProofTheo{LB}
Assume that we have a solution to the bandwidth reservation problem
that requires $o(\lg n)$ time. We will show how to use such a solution
to solve the prefix sum problem in time $o(\log n)$ which contradicts
the lower bound by Fredman (\cite{Fredman82}).

First, we introduce an extra point $n+1$ right to all other points
representing $+\infty$. It is needed since in our problem we are
dealing with open intervals on the right side.
Next, we translate the array of elements in the prefix sum problem
into end-points of intervals. More precisely, the $V(i)$ element of
the array is represented by the interval that starts at point $i$ and
ends at the right most point $n+1$: $\Intvl{i}{n+1}$. Therefore, the reserved
bandwidth at point $p$ is the sum of all reserved bandwidths for
intervals starting at points $j$, where $1 \leq j \leq p$. This gives
us the following translation of prefix-sum problem operations:

\begin{itemize}
\item  the operation $\mathtt{Update}(i,x)$ into
  $\mathtt{Reserve}(x,\Intvl{i}{n+1})$; and
\item  the operation $\mathtt{Retrieve}(j)$ into a query
  $\mathtt{MaxReserved}(\Intvl{j}{j+1})$.
\end{itemize}

This translation gives us an $o(\lg n)$ solution to the prefix sum
problem and hence contradicts the lower bound by Fredman.
\eProof

Note that, the prefix sum as presented by Fredman (\cite{Fredman82})
is also a \emph{static problem} -- i.e. the array of elements neither
expands nor shrinks. On the other hand, the solution we present in the
following section does support insertion of new points (intervals) and
deletion of points (intervals). Hence, by using the translation in the
proof we also get a logarithmic solution to the dynamic version of the
prefix-sum problem.

\section{Upper bound} 

To prove an upper bound we use a data structure called \emph{BinSeT}
that supports the required operations in logarithmic time. Before
going into details of data structure we describe how we represent
reservations.

\subsection{Representation of reservations}
\label{sec:representation}

We do not represent a reservation interval as a single entity, but we
split it into two, what we call, \emph{reservation events}. A
reservation event is a point in time when an increase or decrease in
the amount of a reserved bandwidth occurs. For example, we store a
reservation $R = \{B, \Intvl{t_0}{t_1} \} $ as reservation events
$E_0= (t_0, +B)$ and $E_1= (t_1, -B)$. In other terms, we convert an
interval $\Intvl{t_0}{t_1}$ into two semi-infinite intervals
$\Intvl{t_0}{+\infty}$ and $\Intvl{t_1}{-\infty}$.
Hence, the operations from \rDefn{BRP} are converted:
\begin{itemize}
\item $\mathtt{Reserve}(B, \Intvl{t_0}{t_1} )$ into adding of reservation
  events $E_0=(t_0,+B)$ and $E_1=(t_1,-B)$; and
\item $\mathtt{Free}(B, \Intvl{t_0}{t_1} )$ into adding of reservation
  events $E_0=(t_0,-B)$ and $E_1=(t_1,+B)$; while
\item $\mathtt{MaxReserved}(\Intvl{t_0}{t_1})$, remains the same.
\end{itemize}

If we want to store extra information with each reservation we
introduce an additional dictionary data structure to store this
information and bind the reservation events to records in the
dictionary.

\Subsection(DS){Data Structure}

The binary segment tree BinSeT is a data structure that combines
properties of a binary and a segment tree. The former permits dynamic
insertion and deletion of reservation events and the later answering
queries about the maximum reserved bandwidth. In detail, the leaves
represent and store information about the reservation events, while
each internal node covers a segment (interval) $I=\Intvl{t_0}{t_1}$
and stores information about the values (bandwidth) on that interval.
To ensure $O(\log n)$ worst case performance, we balance BinSeT tree
as an AVL tree (cf.~\cite{Carrano-2001,AVL,Weiss-92}) -- hence we also
need to talk about the height of BinSeT tree.
This gives the following invariance for every node of our data
structure:
\bInvar{invar}
The information stored with the node $n$ representing an interval
$I=\Intvl{t_0}{t_1}$ is the maximum value $\mu_n$ on the interval and
the change $\delta_n$ of the value on the interval.
Besides, with a node is also stored the left-most event in the right
subtree $t_0 < \tau < t_1$. The difference of heights of left
and right subtree is at most one.
\eInvar
Note, if a node covers interval $\Intvl{t_0}{t_1}$, the left subtree
covers interval $\Intvl{t_0}{\tau}$ and the right subtree the interval
$\Intvl{\tau}{t_1}$.

In simpler terms, in the BinSeT tree each node has its own
\emph{local} system of reserved resource values on its interval.  The
system is offset to the global so, that in the beginning of the
interval the value is considered to be $0$. To get total (global)
value of reserved resources one has to add $\delta$-s for all left
siblings on the path from the node to the root.

It is easy to verify the following lemma:
\bLemma{node.mu}
Let $l$ be left child and $r$ right child of an internal node $n$.
Then the equations:
\bEqu{node.mu}
\begin{array}{rcl}
   \delta_n & = & \delta_l + \delta_r
\\ \mu_n    & = & \max( \mu_l, \delta_l + \mu_r)
\end{array}
\eEqu
hold for all nodes $n$.
\eLemma

The detail data structure is represented in \rAlg{ds}.
\bAlg{ds}{pascal}{Binary segment tree definition.} typedef struct \_sBinSeT \{
  tResource       $\mu$;
  tResource       $\delta$;
  tTime           $\tau$;
  unsigned int    height;
  struct \_sBinSeT* left;
  struct \_sBinSeT* right;
\} tBinSeT;
 \eAlg
The structure is slightly different from the one described above since
it does not include times $t_0$ and $t_1$, but only the $\tau$.
However, values $t_0$ and $t_1$ can be implicitly calculated during
recursive descend. At this point we note two things: first,
a node has either two sub-trees (an internal node) or none (a leaf);
and second, a leaf stores in both $\delta$ and $\mu$ the amount of the
reserved bandwidth at the reservation event it represents, and in
$\tau$ the time of the event.  As a consequence of the first
observation we conclude, that the number of internal nodes is one less
than the number of leaves. Since the number of leaves is at most $2n$,
where $n$ is a number of reservation intervals, this proves the
following lemma, under the RAM model:
\bLemma{size}
The size of the BinSeT storing $n$ reservation events is $\Theta(n)$
words.
\eLemma

\subsection{Operations}

Finally we describe how to implement efficiently queries and adding of
reservation events. All our solutions will be recursive and will start
traversing the data structure from the root. We assume that we store
with BinSeT also the time of the first ($t_f$) and the last ($t_l$)
reservation event.  These are also times $t_0$ and $t_1$,
respectively, for the root of the complete BinSeT. If we descend in
the left subtree, then the $t_0$ and $t_1$ for this subtree become
values $t_0$ and $\tau$, respectively, of the current root. We treat
similarly the right subtree. This is also the reason why we need not
store values $t_0$ and $t_1$ with a node.

We start with a query $\mathtt{MaxReserved}$ (see \rAlg{query}).
\bAlg{query}{pascal}{Query \texttt{MaxReserved} in BinSeT.}
tResource MaxReserved(tBinSet* node, tTime t0, tTime t1, tInterval query) \{
  tResource leftMax, rightMax;
  tInterval queryAux;
  if ((t0 == query.t0) \&\& (t1 == query.t1))%
            \CommRight{whole interval -- stopping condition}
    return node->$\mu$;
  if (query.t1 <= node->$\tau$)%
            \CommRight{query in left subinterval}
    return MaxReserved(node->left, t0, node->$\tau$, query);
  if (node->$\tau$ <= query.t0)%
            \CommRight{query in right subinterval}
    return node->left->$\delta$ +
           MaxReserved(node->right, node->$\tau$, t1, query);
  queryAux= query; queryAux.t1= node->$\tau$;%
            \CommRight{query in both subinterval -- so split it}
  leftMax=  MaxReserved(node->left,  t0, node->$\tau$, queryAux);
  queryAux= query; queryAux.t0= node->$\tau$;
  rightMax= MaxReserved(node->right, node->$\tau$, t1, queryAux);
  return max(leftMax, node->left->$\delta$ + rightMax);
\} /* MaxReserved */

\eAlg
Assuming \rInvar{invar} we prove:
\bLemma{query}
The query $\mathtt{MaxReserved}$ in BinSeT takes $O(\log n)$ worst
case time.
\eLemma

\bProofLemma{query}
The correctness of the proof uses induction. Due to the limited
presentation space we give only a justification of the induction step.
Let the query be for the interval $I_q=\Intvl{t_{q0}}{t_{q1}}$ and let the
node cover interval $\Intvl{t_0}{t_1}$. Then we have the following
possibilities:
\begin{itemize}
\item If $t_{q0}=t_0$ and $t_{q1}=t_1$, the answer is exactly $\mu$
  of the node.

\item If $t_{q1} \leq \tau$ then the answer is the same as the answer
  to the same query $I_q$ in the left subtree \Code{left} covering the
  interval $\Intvl{t_0}{\tau}$.

  Similarly, if $\tau \leq t_{q0}$ then the answer is the same as the
  answer to the query $I_q$ in the right subtree \Code{right} covering
  the interval $\Intvl{\tau}{t1}$. However, due to \rLemma{node.mu} we
  have to add left node's $\delta$.

\item Finally, in the most general case when
  $t_0 < t_{q0} < \tau < t_{q1} < t_1$ the answer is because of
  \rLemma{node.mu}
  \[
  \max (\mathtt{MaxReserved}(\Code{left}, \Intvl{t_{q0}}{\tau}),
        \Code{left->}\delta +
        \mathtt{MaxReserved}(\Code{right}, \Intvl{\tau}{t_{q1}} ))
  \enspace .
  \]
\end{itemize}

To see that the running time of the query is logarithmic, i.e.\
proportional to the height of the BinSeT, observe that the third case
occurs only once, while the tree is balanced in the AVL-sense.
\eProof

The last operation is \Code{Add} that adds a reservation event.
Note, that we \emph{never} explicitly delete a reservation event, we
might just add a reservation event with a negative value (see section
\ref{sec:representation}).
\bLemma{update}
Adding of a reservation event into BinSeT can be done in
$O(\log n)$ worst case time.
\eLemma

\bProofLemma{update}
Let us assume that we are adding a reservation event at time $t_r$ and
for the value $B_r$.
\bAlg{add}{pascal}{Adding of a reservation event in BinSeT.}
tBinSet* Add(tBinSet* node, tTime $t_r$, tResource $B_r$) \{

  if (node->left != null) \{         \CommRight{\textsc{We are not at the leaf yet.}}
    if ($t_r$ < node->$\tau$) \{
      node->left= Add(node->left, $t_r$, $B_r$);
      if (node->left == null) \{         \CommRight{we lost the leaf}
        free(node); return node->right;  \CommRight{but we need no rebalancing}
      \}
    \} else \{ ... \}               \CommRight{similarly for the right subtree}
    node->$\delta$+= $B_r$; \CommRight{update $\delta$ and $\mu$ -- see \rEqu{node.mu}}
    node->$\mu$= max(node->left->$\mu$, node->left->$\delta$ + node->right->$\mu$);
    node= Rebalance(node);
    return node;
  \}
                              \CommRight{\textsc{We are at the leaf.}}
  if ($t_r$ != node->$\tau$)   return Insert(node, $t_r$, $B_r$);
  else \{
    node->$\mu$= node->$\delta$= node->$\mu$ + $B_r$;
    if (node->$\mu$ != 0) return node;
    else \{ free(node); return null; \}
  \}
\} /* Add */

\eAlg
We start (see \rAlg{add}) at the root and recursively descend to the
leaves.  The decision into which subtree to descend is based on the
node's value $\tau$ and $t_r$:
when $t_r < \tau$, we descend into the left subtree and otherwise into
the right one. Note, we always go all the way to the leaves.

The time $\tau$ of the reached leaf can be either the same as $t_r$ or
not.  If it is not, we create a new internal node \Code{newNode} and
make the reached leaf one of its leaves. Besides, we create a new leaf
with an added reservation event and properly update the values. For
details see \rAlg{insert}.
\bAlg{insert}{pascal}{Insertion of a new reservation event in BinSeT.}
tBinSet* Insert(tBinSet* oldLeaf, tTime $t_r$, tResource $B_r$) \{
  tBinSet* newLeaf;
  tBinSet* newNode;
            \CommRight{First make a new leaf out of an inserted event:}
  newLeaf= (tBinSet*) malloc( sizeof(tBinSet) );
  newLeaf->$\mu$= newLeaf->$\delta$= $B_r$;%
            \CommRight{set first as a segment tree}
  newLeaf->$\tau$= $t_r$;
  newLeaf->height= 1;%
            \CommRight{and then as a binary tree.}
  newLeaf->left= newLeaf->right= null;
            \CommRight{And then make a new internal node:}
  newNode= (tBinSet*) malloc( sizeof(tBinSet) );
  newLeaf->height= 2;%
            \CommRight{now first set as a binary tree}
  if (oldLeaf->$\tau$ < $t_r$) \{ newNode->left= oldLeaf; newLeaf->right= newLeaf; \}
  else \{ newNode->left= newLeaf; newLeaf->right= oldLeaf; \}
  newNode->$\delta$= newNode->left->$\delta$ + newNode->right->$\delta$;%
\CommRight{and then as a segment tree}
  newNode->$\mu$= max(newNode->left->$\mu$,
                  newNode->left->$\delta$ + newNode->right->$\mu$);
  return newNode;
\} /* Insert */

\eAlg

On the other hand, if $\tau = t_r$ we add value $B_r$ to leaf's values
$\mu$ and $\delta$. If new values are not $0$ we are done. However, if
they are $0$ we have to delete the leaf and replace its parent with
leaf's sibling. We also delete the parent. On the way back to the root
we update $\delta$-s and $\mu$-s as required in \rEqu{node.mu}.
\rAlg{add} gives a skeleton of the algorithm.

It remains to describe the rebalancing of BinSeT (see call of
\Code{Rebalance} function in \rAlg{add}). Since BinSeT is an AVL-like
tree, we rebalance it using regular single and double rotations. While
the details of when and how to perform the rotations are explained in
most textbooks (cf.\ \cite{Carrano-2001,Weiss-92}) we concentrate only
on updates of values $\mu$ and $\delta$.  Observe that the value
$\tau$ does not change during rotations.

First consider a single rotation shown in \rFig{single} (we are
omitting description of a mirroring single rotation).
\fig{3cm}{single}{Single rotation.}
The new values of nodes $b$ and $d$, they are marked with a prime sign,
are computed using the formulae:
\bEqu{single}
  \begin{array}{rcl}
    d.\delta' &=& b.\delta
\\  b.\delta' &=& b.\delta - E.\delta
  \end{array}
\hfill
\hspace{0.1\textwidth}
\hfill
  \begin{array}{rcll}
    d.\mu' &=& b.\mu
\\  b.\mu' &=& \max(A.\mu, A.\delta + C.\mu) & \textrm{by \rEqu{node.mu}}
  \end{array}
\eEqu
Observe, that the order in which new values are computed is important:
therefore we first compute $\delta$ and $\mu$ values at $d$ and
afterwards at $b$.

Similarly we compute new values in double rotation (cf.\
\rFig{double}):
\fig{3cm}{double}{Double rotation.}
\bEqu{double}
  \begin{array}{rcl}
    d.\delta' &=& b.\delta
\\  f.\delta' &=& f.\delta - C.\delta
\\  b.\delta' &=& f.\delta'
  \end{array}
\hfill
\hspace{0.1\textwidth}
\hfill
  \begin{array}{rcll}
    d.\mu' &=& b.\mu
\\  b.\mu' &=& \max(A.\mu, A.\delta + C.\mu) & \textrm{by \rEqu{node.mu}}
\\  f.\mu' &=& \max(E.\mu, E.\delta + G.\mu) & \textrm{by \rEqu{node.mu}}
  \end{array}
\eEqu

To prove the correctness of \rAlg{add} we need to see that it
preserves \rInvar{invar}. First, if a new reservation point is added
in the interval the $\delta$ should be changed exactly for this value.
This is done in line 9 of \rAlg{add}. In the following line new $\mu$
is computed according to \rEqu{node.mu} and hence also this part of
invariance is kept.

Finally, the rebalancing keeps the difference in heights between the
left and right subtrees always at most one. Consequently, the height
of the tree is $O(\log n)$ and the running time of \rAlg{add} is also
$O(\log n)$.
\eProof

Our data structure uses AVL-like balancing technique, but it could use
any one. For more details on balancing and balance binary trees see
\cite{CormenLeissersonRivest01Introduction,Andersson1990} or any other
text book.

This brings us to the final theorem:
\bTheo{BRP}
The Bandwidth Reservation Problem can be solved under the comparison
based machine model in $\Theta(\log n)$ time per operation and in
$\Theta(n)$ words of space. This is tight.
\eTheo

Obviously it is straight forward to adapt the solution to handle also
queries of the minimum reserved bandwidth. Moreover, using the
translation in \rTheo{LB} we also get a logarithmic time solution to
the dynamic versions of partial sum problem and of prefix sum problem.

A practical improvement is to store with a node not its $\delta$ and
$\mu$, but rather its children's $\mu$-s and left child's $\delta$
(the right child's $\delta$ is actually never used!). Using this
information it is easy to compute also node's $\mu$ using
\rEqu{node.mu}. One would think that the size of the data structure
increases after such a modification. But it does not, since we do not
need leaves at all. Moreover, since in \rAlg{add}, and in
\rEqu{single} and \rEqu{double} we no more need to access children,
everything runs faster because of fewer cache misses.

\section{Conclusions} 

We showed that the data structure BinSeT (binary segment tree) solves
the dynamic version of the Bandwidth Reservation Problem optimally
(space- and time-wise) under the comparison based model.
The solution requires $\Theta(\log n)$ time for the queries and
updates and $\Theta(n)$ space.
It substantially improves solution presented in \cite{schelen99} which
restricted the maximum allowed reservation intervals and their
smallest granularity.

Using BinSeT we also solve dynamic versions of prefix sum and partial
sum problems. Interesting enough, asymptotically the dynamic solution
has the same time and space complexity as the static version.

There are a number of open problems left. For example, what are lower
and upper bounds under the cell probe model and bounded universe?
Interesting question is also whether can we benefit from the fact that
time always increases? At least on the average?

\bibliography{BinSeT} 

\appendix 

\section{An Example}

Bottom of \rFig{BinSeT2-1} gives an example of a reservations made
during 16 time slots.
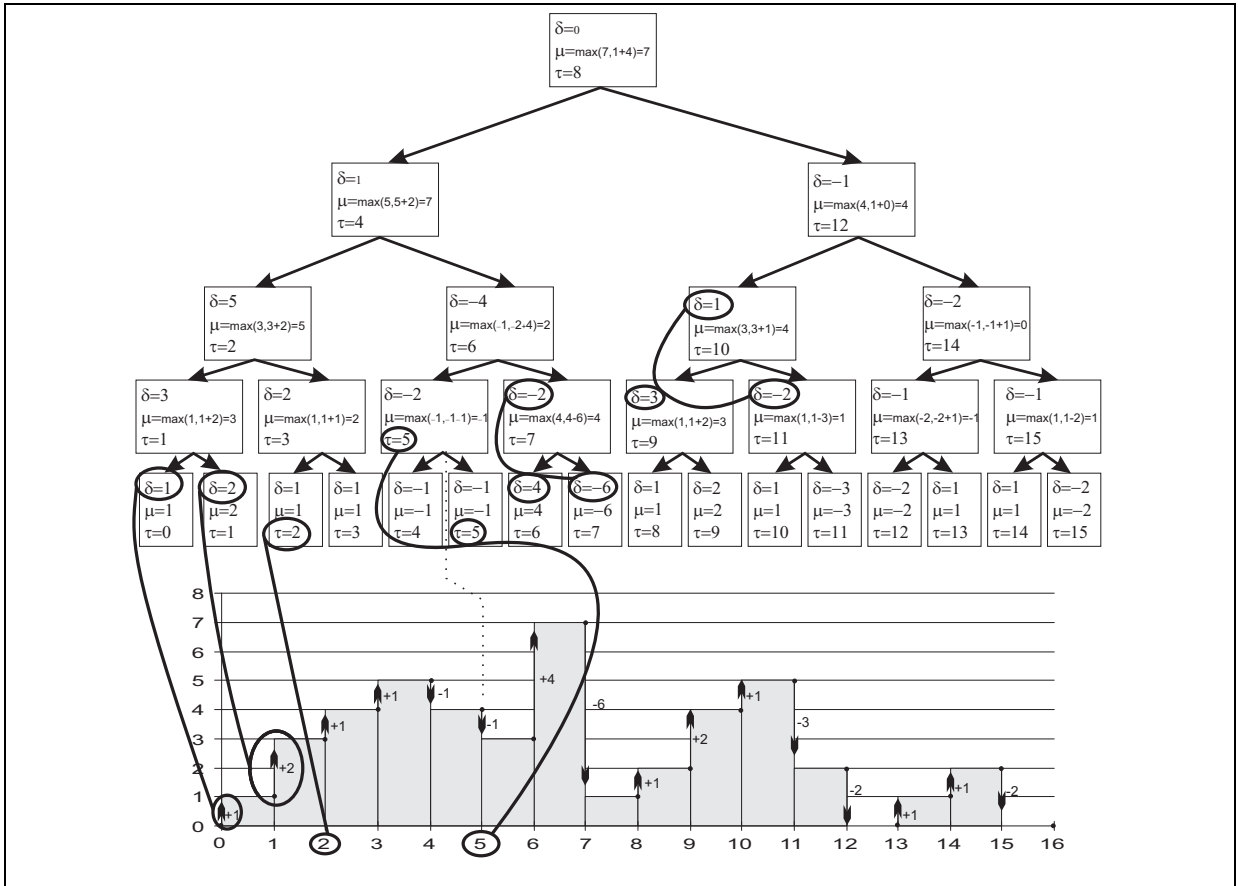
\fig{3cm}{BinSeT2-1}{Example of a BinSeT tree.}
In the upper part of the figure is presented a BinSeT tree as build
over the presented reservations. Additional arrows explain how
particular values of $\delta$, $\mu$ and $\tau$ are computed from the
reservations.

In \rFig{BinSeT2-2} is shown a detail from the example.
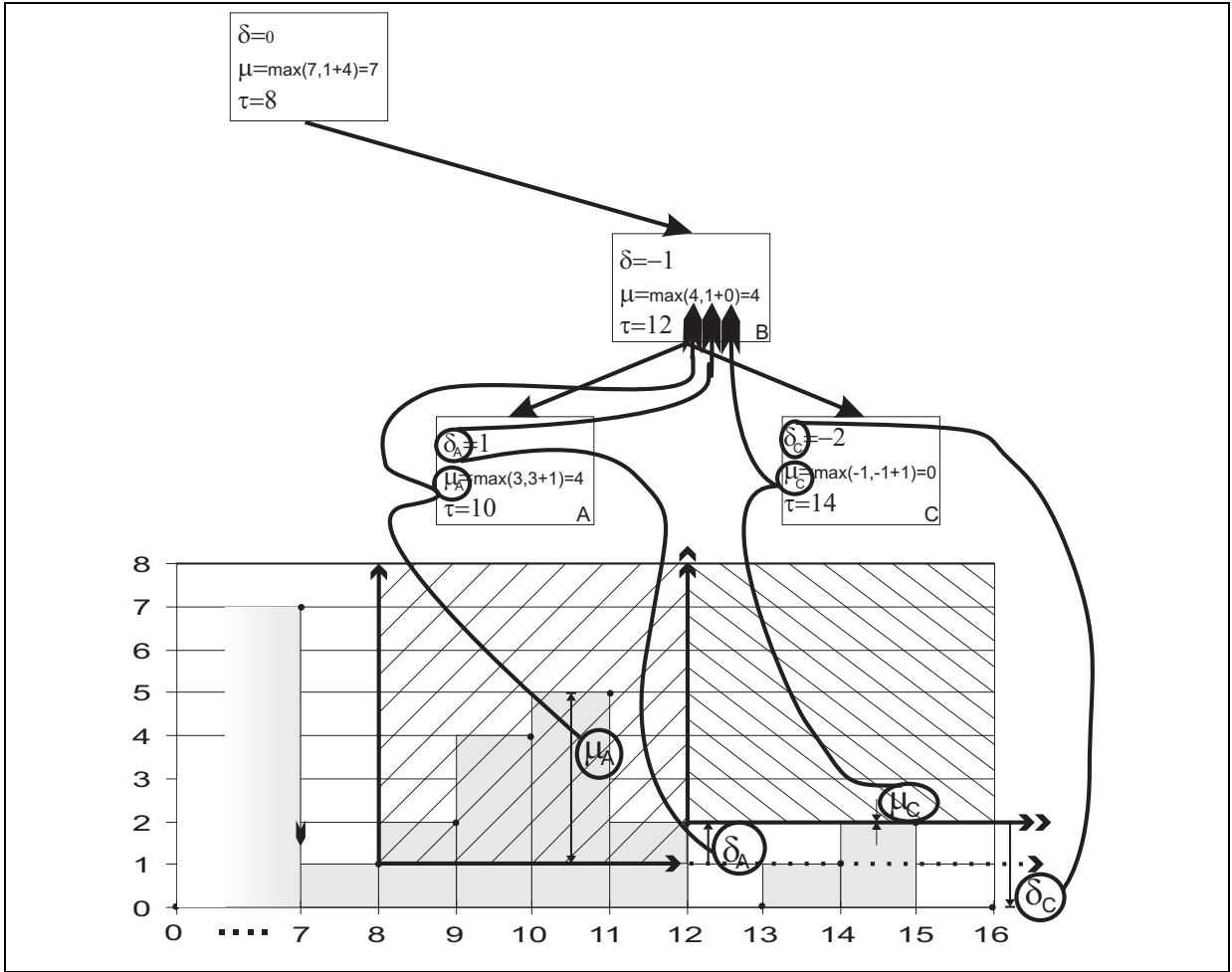
\fig{3cm}{BinSeT2-2}{Local systems of internal nodes $A$ and $C$.}
It presents ``local systems'' mentioned in \RefPart{DS} for internal
nodes $A$ and $C$. The systems are presented with two different
patterns: the first is expanding over slots 8 to 12 and the second one
from 12 to 16. The figure also depicts $\delta$ and $\mu$ values for
both nodes.

\end{document}